\documentclass[]{iopart}
\usepackage{graphicx}
\usepackage{iopams}

\newcommand{\half}{{\case{1}{2}}}
\newcommand{\fourth}{{\case{1}{4}}}

\begin{document}
\vspace{-2.5cm} 
\title[Generalized Harmonic Evolution System]
{A New Generalized Harmonic Evolution System}

\author{Lee Lindblom${}^1$, Mark A. Scheel${}^1$, Lawrence E. Kidder${}^2$,\\
Robert Owen${}^1$, and Oliver Rinne${}^1$}
\address{${}^1$ Theoretical Astrophysics 130-33, California Institute of
Technology, Pasadena, CA 91125}
\address{${}^2$ Center for Radiophysics and Space Research,
Cornell University, Ithaca, New York, 14853}

\date{\today}

\begin{abstract}
A new representation of the Einstein evolution equations is presented
that is first order, linearly degenerate, and symmetric hyperbolic. This
new system uses the generalized harmonic method to specify the coordinates,
and exponentially suppresses all small short-wavelength constraint
violations.  Physical and constraint-preserving boundary conditions
are derived for this system, and numerical tests that demonstrate the
effectiveness of the constraint suppression properties and the
constraint-preserving boundary conditions are presented.
\end{abstract}


\submitto{\CQG}

%%%%%%%%%%%%%%%%%%%%%%%%%%%%%%%%%%%%%%%%%%%%%%%%%%%%%%%%%%%%%%%%%%%%%%%%%%%%%%%
\section{Introduction}
\label{s:Introduction}

Harmonic and generalized harmonic (GH) coordinates have played
important roles in general relativity theory from the very beginning.
Einstein used harmonic (then called isothermal) coordinates in his
analysis of candidate theories of gravitation (as recorded in his
Zurich notebook of 1912) before general relativity even
existed~\cite{Sauer1998}, DeDonder
used them to analyze the characteristic structure of general relativity 
in 1921~\cite{deDonder1921,deDonder1927}, and Fock used them to
analyze gravitational waves in 1955~\cite{Fock1959}.  
Harmonic coordinates played an important
role in the proofs of the well-posedness of the Cauchy problem for the
Einstein equations by Choquet-Bruhat in
1952~\cite{Choquet1952,Choquet1962} and by Fischer and Marsden in
1972~\cite{fischer_marsden72}.  
Harmonic coordinates have also been used to obtain numerical solutions
of Einstein's equations by Garfinkle~\cite{Garfinkle2002} and by Winicour 
and collaborators~\cite{Szilagyi2002,Szilagyi2003,Babiuc2006}.
The idea of specifying {\it arbitrary}
coordinate systems using a generalization of harmonic coordinates was
introduced by Friedrich in 1985~\cite{Friedrich1985}.  And quite
recently the GH approach to specifying coordinates played an
important, perhaps seminal, role in the state-of-the-art numerical
simulations of the final inspiral and merger of binary black-hole
systems by Pretorius~\cite{Pretorius2005c, Pretorius2005a} using a
form of the equations suggested by Gundlach, et
al.~\cite{Gundlach2005}.

We think there are two important properties that have made harmonic or
GH coordinates such an important tool throughout the history of
general relativity theory.  The first property is well known: this
method of specifying the coordinates transforms the principal parts of
the Einstein equations into a manifestly hyperbolic form, in which
each component of the metric is acted on by the standard second-order
wave operator.  The second property is not as widely appreciated: this
method of specifying coordinates fundamentally transforms the
constraints of the theory.  This new form of the constraints makes it
possible to modify the evolution equations in a way that prevents
small constraint violations from growing during numerical
evolutions---without changing the physical solutions of the system and
without changing the fundamental hyperbolic structure of the
equations.  The purpose of this paper is to explore and understand
these important properties and to extend the GH evolution system in a
way that makes it even more useful for numerical computations.  In
Sec.~\ref{s:GeneralizedHarmonic} we review the modified form of the GH
evolution system of Gundlach, et al. and Pretorius.  We convert and
extend this system in Sec.~\ref{s:FirstOrderSystem} into a
symmetric-hyperbolic first-order evolution system that has constraint
suppression properties comparable to those of the second-order
system.  We derive and analyze the well-posedness of constraint-preserving 
and physical boundary conditions for this new first-order
system in Sec.~\ref{s:BoundaryConditions}, and in
Sec.~\ref{s:NumericalResults} we present numerical tests that
demonstrate the effectiveness of its constraint suppression properties
and the new constraint-preserving boundary conditions.

%%%%%%%%%%%%%%%%%%%%%%%%%%%%%%%%%%%%%%%%%%%%%%%%%%%%%%%%%%%%%%%%%%%%%%%%%%%%%%%
\section{Generalized Harmonic Evolution System}
\label{s:GeneralizedHarmonic}

Harmonic (sometimes called wave) coordinates are functions $x^a$ that
satisfy the covariant scalar wave equation.  These coordinates are
very useful because they significantly simplify the second-derivative
terms in the Ricci curvature tensor.  To see this explicitly, consider
a spacetime with metric tensor $\psi_{ab}$:
\begin{eqnarray} 
ds^2 &=& \psi_{ab}dx^a dx^b. 
\end{eqnarray}
(We use Latin indices from the first part of the alphabet $a, b, c,
...$ to denote $4$-dimensional spacetime quantities.)  A coordinate
$x^b$ is called harmonic if it satisfies the scalar wave equation,
\begin{eqnarray}
0=\psi_{ab}\nabla^c\nabla_c x^b = -\Gamma_a,\label{e:HarmonicDef}
\end{eqnarray}
where $\nabla_c$ denotes the covariant derivative compatible with
$\psi_{ab}$, and $\Gamma_a\equiv \psi^{bc}\Gamma_{abc}$ is the trace
of the standard Christoffel symbol $\Gamma_{abc}$:
\begin{eqnarray}
\Gamma_{abc}&=&\half
(\partial_b\psi_{ac}
+\partial_c\psi_{ab}
-\partial_a\psi_{bc}).\label{e:Christoffel}
\end{eqnarray}
The right side of Eq.~(\ref{e:HarmonicDef}) is just the expression for
this covariant wave operator acting on $x^b$ in terms of partial
derivatives and Christoffel symbols.

The Ricci curvature tensor can be written as 
\begin{eqnarray}
R_{ab} &=& -\half \psi^{cd}\partial_c\partial_d 
\psi_{ab} + \nabla_{(a}\Gamma_{b)}
%\nonumber\\&&
+\psi^{cd}\psi^{ef}\bigl(
\partial_e\psi_{ca}\partial_f\psi_{db}
-\Gamma_{ace}\Gamma_{bdf}\bigr),\nonumber\\
\label{e:RicciDef}
\end{eqnarray}
in any coordinate system, where $\nabla_a\Gamma_b\equiv \partial_a
\Gamma_b - \psi^{cd}\Gamma_{cab}\Gamma_d$.  In harmonic coordinates,
$\Gamma_a=0$, so the only second-derivative term remaining in the
Ricci tensor is $\psi^{cd}\partial_c\partial_d\psi_{ab}$.  Therefore
in harmonic coordinates the vacuum Einstein equations, $R_{ab}=0$, 
form a manifestly hyperbolic system~\cite{Choquet1952},
\begin{eqnarray}
\psi^{cd}\partial_c\partial_d \psi_{ab} &=& 2\,
\psi^{cd}\psi^{ef}\bigl( \partial_e\psi_{ca}\partial_f\psi_{d b}
-\Gamma_{ace}\Gamma_{bdf}\bigr).
\label{e:VacuumHarmonic}
\end{eqnarray}

Friedrich~\cite{Friedrich1985} (and independently
Garfinkle~\cite{Garfinkle2002}) realized that the manifestly
hyperbolic form of the Einstein system, Eq.~(\ref{e:VacuumHarmonic}),
can also be achieved for {\it arbitrary} coordinates, if the choice of
coordinates is fixed in a certain (but non-standard) way.  This
alternate method of specifying the choice of coordinates, which we call the
generalized harmonic (GH) method, is implemented by assuming that the
coordinates satisfy the inhomogeneous wave equation,
\begin{eqnarray}
H_a(x,\psi)=\psi_{ab}\nabla_c\nabla^c x^b = -\Gamma_a,
\label{e:GHCond}
\end{eqnarray}
where $H_{a}(x,\psi)$ is an arbitrary but fixed algebraic function of
the coordinates $x^a$ and the metric $\psi_{ab}$ (but not its
derivatives).  In these GH coordinates $H_a=-\Gamma_a$, so the vacuum
Einstein equations are again manifestly hyperbolic:
\begin{eqnarray}
\psi^{cd}\partial_c\partial_d 
\psi_{ab}  &=& - 2\nabla{}_{(a}{H}{}_{b)}
+2\,\psi^{cd}\psi^{ef}\bigl(
\partial_e\psi_{ca}\partial_f\psi_{d b}
-\Gamma_{ace}\Gamma_{bdf}\bigr)
.\label{e:GHSystem}
\end{eqnarray}
The term containing $H_{b}$ on the right side of
Eq.~(\ref{e:GHSystem}) is a pre-specified algebraic function (of
$x^a$ and $\psi_{ab}$) that operates as a source term, rather
than one of the principal terms containing second derivatives of
$\psi_{ab}$.  The principal (i.e., second-derivative) parts of this GH
evolution system, Eq.~(\ref{e:GHSystem}), are therefore identical to
those of the harmonic evolution system, Eq.~(\ref{e:VacuumHarmonic}).

To understand the GH method of specifying coordinates more
clearly, it is helpful to compare it to the more traditional way of
specifying coordinates with the lapse and the shift.  To do this we
introduce a foliation of the spacetime by spacelike hypersurfaces, and
adopt a coordinate system, $\{t,x^k\}$, with the $t=$ constant
surfaces being the leaves of this foliation.  The traditional lapse
$N$, shift $N^k$, and $3$-dimensional spatial metric $g_{ij}$
associated with this coordinate system are then defined by
\begin{eqnarray}
ds^2 &=& \psi_{ab}dx^a dx^b =-N^2 dt^2 + g_{ij}(dx^i+ N^i dt)(dx^j+N^j
dt).\label{e:ADMMetric}
\end{eqnarray}
(We use Latin indices $i, j, k, ...$ to denote 3-dimensional spatial
quantities; while Latin indices from the first part of the alphabet
$a,b,c, ...$ will continue to denote 4-dimensional quantities.)
Expressing the GH coordinate condition, Eq.~(\ref{e:GHCond}), in this
3+1 language implies evolution equations for the lapse and shift:
\begin{eqnarray}
\partial_t N - N^k\partial_k N &=&
-N\bigl(H_t - N^i H_i + N K\bigr),\\    
\partial_t N^i - N^k\partial_k N^i &=& N g^{ij}\Bigl[
N \bigl(H_j+g^{kl}\Gamma_{jkl}\bigr) -\partial_j N\Bigr],
\end{eqnarray}
where $K$ is the trace of the extrinsic curvature.  Specifying the GH
gauge function $H_a(x,\psi)$ therefore determines the time derivatives
of the lapse $N$ and shift $N^k$, and hence the evolution of the gauge
degrees of freedom of the system.  Some gauge conditions (e.g., $N=1$,
$N^k=0$) may not be simple conditions on $H_a$, just as some gauge
conditions (e.g., $H_a=0$) are not simple conditions on $N$ and $N^k$.
In this paper we restrict attention to the cases where $H_a(x,\psi)$
is a specified algebraic function.  Any chosen coordinates can clearly
be described ({\it ex post facto}) by an $H_a$ of this form.
But $H_a$ may also be specified in more general ways, e.g., by giving
evolution equations for $H_a$~\cite{Pretorius2005c}.  We expect (but
have not proven) that any coordinates can be obtained by specifying
{\it a priori\/} suitable (possibly complicated) conditions on $H_a$.
%%%%%%%%%%%%%%%%%%%%%%%%%%%%%%%%%%%%%%%%%%%%%%%%%%%%%%%%%%%%%%%%%%%%%%%%%%%%%%%
\subsection{Constraint Evolution}
\label{s:ConstraintEvolution}

Our experience in solving the Einstein equations numerically is that
small constraint violations typically grow into large constraint
violations that quickly make the solutions unphysical.  We think it is
essential therefore to understand the constraints and how violations
of those constraints evolve with time.  To this end it is helpful to
consider the following representation of the GH
system, Eq.~(\ref{e:GHSystem}):
\begin{eqnarray}
0=R_{ab}-\nabla_{(a}{\cal C}_{b)}
,\label{e:GHEq2}
\end{eqnarray}
where $R_{ab}$ is the Ricci tensor defined in Eq.~(\ref{e:RicciDef}),
and ${\cal C}_a$ is defined as 
\begin{eqnarray}
{\cal C}_a = {H}_a +\Gamma_a.
\label{eq:GaugeConstraint}
\end{eqnarray}
From this perspective the condition ${\cal C}_a=0$ serves as the
constraint that ensures the coordinates satisfy the GH coordinate
condition, Eq.~(\ref{e:GHCond}).  It is straightforward to verify that
Eq.~(\ref{e:GHEq2}) is equivalent to the GH evolution equations,
Eq.~(\ref{e:GHSystem}).  This form of the GH system,
Eq.~(\ref{e:GHEq2}), is also formally equivalent to the Z4
system~\cite{Bona2003} (in
the sense that there is a one-to-one correspondence between solutions
of the two systems), where the constraint ${\cal C}_a$ plays the role
of the Z4 vector field~\cite{Gundlach2005}.  The systems differ however
in the way the fields are evolved: in the Z4 system the
field ${\cal C}_a$ is evolved as a separate dynamical field,
while in the GH representation ${\cal C}_a$ is treated 
as a constraint which is not evolved separately.

The evolution equation for the constraints is easily deduced from the
GH evolution system, Eq.~(\ref{e:GHEq2}): take the
divergence of the trace-reversed  Eq.~(\ref{e:GHEq2}), use the
contracted Bianchi identity $\nabla^a R_{ab}-\half\nabla_b R=0$, and
exchange the order of covariant derivatives with the Ricci identity,
yielding
\begin{eqnarray}
0=\nabla^b\nabla_b {\cal C}_a + R_{ab}\,
{\cal C}^{\,b}.\label{e:ConstraintEvol0}
\end{eqnarray}
Finally the Ricci tensor can be eliminated using Eq.~(\ref{e:GHEq2})
to produce the following equation for the evolution of the
constraints~\cite{Friedrich2005}:
\begin{eqnarray}
0=\nabla^b\nabla_b {\cal C}_a + {\cal C}^{\,b}
\nabla{}_{(a}{\cal C}{}_{b)}.\label{e:ConstraintEvol}
\end{eqnarray}
This equation guarantees that the constraints ${\cal C}_a$ will remain
zero within the domain of dependence of an initial surface on which
${\cal C}_a=\partial_t{\cal C}_a=0$.  Thus the GH
evolution system is self-consistent.

The standard Hamiltonian and momentum constraints of general
relativity are encoded in the constraints of the GH
system in an interesting way.  Let $t^a$ denote the unit timelike
normal to the $t=$ constant surfaces of the foliation used in
Eq.~(\ref{e:ADMMetric}).  The standard Hamiltonian and momentum
constraints are combined here into the single $4$-dimensional
momentum constraint ${\cal M}_a$, which is given by the contraction of
$t^a$ with the Einstein curvature tensor:
\begin{eqnarray}
{\cal M}_a \equiv \bigl(R_{ab}-\half \psi_{ab}R\bigr)t^b.
\end{eqnarray}
Using Eq.~(\ref{e:GHEq2}) for a spacetime that satisfies the
GH evolution system, we see that
\begin{eqnarray}
t^b \nabla_b{\cal C}_a = 2 {\cal M}_a 
+ (g^{bc}t_a 
- t^c g^b{}_a) \nabla_b{\cal C}_c,\label{e:ConstraintDot}
\end{eqnarray}
where $g_{ab}= \psi_{ab}+t_a t_b$ is the intrinsic metric to the
$t=$ constant hypersurfaces.  Specifying the initial data needed to
determine the evolution of the constraints, $\{{\cal
C}_a,\partial_t{\cal C}_a\}$, via Eq.~(\ref{e:ConstraintEvol}) is
equivalent therefore to specifying the more usual representation of
the constraints, $\{{\cal C}_a,{\cal M}_a\}$, on that surface.  

%%%%%%%%%%%%%%%%%%%%%%%%%%%%%%%%%%%%%%%%%%%%%%%%%%%%%%%%%%%%%%%%%%%%%%%%%%%%%%%
\subsection{Constraint Damping}
\label{s:ConstraintDamping}

The impressive numerical simulations of binary black-hole spacetimes
performed recently by Pretorius~\cite{Pretorius2005c, Pretorius2005a}
are based on a modified form of the GH evolution system suggested by
Gundlach, et al.~\cite{Gundlach2005}.  This modified system has the
remarkable property that it causes constraint violations to be damped
out as the system evolves. The modified system is given by
\begin{eqnarray}
0 = R_{ab}-\nabla{}_{(a}{\cal C}{}_{b)}
+\gamma_0\bigl[ t{}_{(a}{\cal C}{}_{\,b)}-\half\psi_{ab}\,
t^c {\cal C}_c\bigr],\label{e:EDefNew}
\end{eqnarray}
where $t^a$ (as before) is the future directed timelike unit normal to
the $t=$ constant surfaces, and $\gamma_0$ is a constant that
determines the timescale on which the constraints are damped.  This
system can also be written more explicitly as
\begin{eqnarray}
\psi^{cd}\partial_c\partial_d 
\psi_{ab}  &=&- 2\nabla{}_{(a}{H}{}_{b)}
+\,2\,\psi^{cd}\psi^{ef}\bigl(
\partial_e\psi_{ca}\partial_f\psi_{d b}
-\Gamma_{ace}\Gamma_{bdf}\bigr)\nonumber\\
&&\quad+\,\gamma_0 \bigl[2\delta^c{}_{\!(a}t{}_{b)}-\psi_{ab}\,
t^c\bigr] ({H}_c+\Gamma_c).\label{e:PretoriusSystem}
\end{eqnarray}
This system is manifestly hyperbolic since the additional constraint
damping terms (i.e., those proportional to $\gamma_0$) do not modify
the principal parts of the standard GH evolution
system.  It is also clear that the constraint-satisfying solutions of
this system are identical to those of the standard Einstein system.

In order to understand how this modification affects the constraints,
we must analyze the associated constraint evolution system.  This can
be done by following the same steps that lead to
Eq.~(\ref{e:ConstraintEvol0}), but in this case we obtain
\begin{eqnarray}
0=\nabla^b\nabla_b {\cal C}_a + R_{ab}\,
{\cal C}^{\,b}-2\gamma_0 \nabla^b [\,t{}_{(b}{\cal C}{}_{a)}],
\end{eqnarray}
or using Eq.~(\ref{e:EDefNew}),
\begin{eqnarray}
0&=&\nabla^c\nabla_c {\cal C}_a 
-2\gamma_0 \nabla^b [\,t_{(b}{\cal C}_{a)}]
+ {\cal C}^{\,b}\nabla_{(a}{\cal C}_{b)}
-\half\gamma_0 \,t_a {\cal C}^{\,b}{\cal C}_{b}.\label{e:CDamping}
\end{eqnarray}
This constraint evolution system has the same principal part as the
unmodified system, Eq.~(\ref{e:ConstraintEvol}). Therefore the same
arguments about the self-consistency of the system and the
preservation of the constraints within the domain of dependence apply.
Similarly the relationship between the ${\cal C}_a$ constraint and the
standard $4$-dimensional momentum constraint is not changed in any
essential way: setting ${\cal C}_a = \partial_t {\cal C}_a =0$ on a
$t=$constant surface is still equivalent to setting ${\cal C}_a =
{\cal M}_a =0$ there.

Consider the properties of the constraint evolution system for states
that are very close to the constraint-satisfying submanifold ${\cal
C}_a=\partial_t{\cal C}_a=0$.  We can ignore the terms in
Eq.~(\ref{e:CDamping}) that are quadratic in ${\cal C}_a$ in this case,
so the constraint evolution system reduces to
\begin{eqnarray}
0&=&\nabla^b\nabla_b {\cal C}_a 
-2\gamma_0 \nabla^b [\,t_{(b}{\cal C}_{a)}].
\end{eqnarray}
Gundlach, et al.~\cite{Gundlach2005} have shown that all the ``short
wavelength'' solutions to this constraint evolution system are damped
at either the rate $e^{-\gamma_0t}$ or $e^{-\gamma_0t/2}$.  This
explains how the addition of the terms proportional to $\gamma_0$ in
the modified GH system, Eq.~(\ref{e:EDefNew}), tend to damp out small
constraint violations.  This also explains (in part) why the numerical
evolutions of this system by Pretorius were so successful.  A complete
understanding of how the long wavelength constraints are damped (or
not) in generic spacetimes would also be quite interesting, but this
is not yet fully understood.

%%%%%%%%%%%%%%%%%%%%%%%%%%%%%%%%%%%%%%%%%%%%%%%%%%%%%%%%%%%%%%%%%%%%%%%%%%%%%%%
\section{New First-Order GH Evolution System}
\label{s:FirstOrderSystem}

In this section we present a new first-order representation of the
modified GH evolution system, which will (we think) be a useful
counterpart to the second-order system described in
Sec.~\ref{s:ConstraintDamping} above.  There is an extensive
mathematical literature on first-order evolution systems that
clarifies numerous issues of great importance in numerical relativity,
e.g., how to formulate well-posed boundary conditions~\cite{Rauch1985,
Secchi1996a, Secchi1996b}, which systems form shocks~\cite{liu1979},
etc. We have also been more successful implementing first-order
systems in our spectral evolution code.

The principal part of each component of the modified GH system,
$\psi^{cd} \partial_c \partial_d \psi_{ab}$, is the same as the
principal part of the covariant scalar-field system.  So a first-order
representation of the GH system can be constructed simply by adopting
the methods used for scalar fields~\cite{Scheel2004, Holst2004}.
Using this method and the usual 3+1 coordinates,
Eq.~(\ref{e:ADMMetric}), a first-order representation of the GH system
can be written down, and indeed was written down (in essentially this
form) by Alvi~\cite{Alvi2002}:
\begin{eqnarray}
\partial_t\psi_{ab}-N^k\partial_k\psi_{ab} &\simeq& 0,
\label{e:AlviPsiDot}\\
\partial_t\Pi_{ab} - N^k\partial_k\Pi_{ab} 
+ N g^{ki}\partial_k\Phi_{iab} &\simeq&0,
\label{e:AlviPiDot}\\ 
\partial_t\Phi_{iab}-N^k\partial_k\Phi_{iab}
+N\partial_i\Pi_{ab}&\simeq&0,
\label{e:AlviPhiDot}
\end{eqnarray}
where $\Phi_{iab}=\partial_i\psi_{ab}$ and
$\Pi_{ab}=-t^c\partial_c\psi_{ab}$ are new fields introduced to
represent the first derivatives of $\psi_{ab}$.  The notation $\simeq$
indicates that only the principal parts of the equations (i.e., the
parts containing derivatives of the fields) are displayed.

In the discussion that follows, it will be helpful to discuss
first-order evolution systems like this using a more compact and more
abstract notation.  Thus, we let $u^\alpha = \{\psi_{ab}, \Pi_{ab},
\Phi_{iab}\}$ denote the collection of dynamical fields; and the
evolution system for these fields can be written as
\begin{eqnarray}
\partial_t u^\alpha + A^{k\,\alpha}{}_\beta \partial_k u^\beta = F^\alpha,
\end{eqnarray}
where $A^{k\,\alpha}{}_\beta$ and $F^\alpha$ may depend on $u^\alpha$
but not its derivatives.  We use Greek indices throughout this paper
to label the collection of dynamical fields.  The principal part of
this system is written abstractly as $\partial_t u^\alpha +
A^{k\,\alpha}{}_\beta \partial_k u^\beta \simeq 0$, so
Eqs.~(\ref{e:AlviPsiDot})--(\ref{e:AlviPhiDot}) determine the matrix
$A^{k\,\alpha}{}_\beta$ but not $F^\alpha$ for this system.  First
order evolution systems of this form are called {\it symmetric
hyperbolic} if there exists a symmetric positive definite matrix
$S_{\alpha\beta}$ (the symmetrizer) on the space of fields that
satisfies the condition $S_{\alpha\mu} A^{k\,\mu}{}_\beta \equiv
A^k_{\alpha\beta}=A^k_{\beta\alpha}$.  The mathematical literature on
symmetric hyperbolic systems is extensive, and includes for example
strong existence and uniqueness theorems~\cite{fischer_marsden72, 
Rauch1985, Secchi1996a,
Secchi1996b}.  Alvi's representation of the GH system~\cite{Alvi2002}
is symmetric hyperbolic, as was a similar representation of the
Einstein system (for the case of harmonic coordinates) given earlier
by Fischer and Marsden~\cite{fischer_marsden72}.

Alvi's first-order representation of the GH system
has two serious problems: First, the use of the field
$\Phi_{iab}$ introduces a new constraint,
\begin{eqnarray}
{\cal C}_{iab} =\partial_i\psi_{ab} -
\Phi_{iab},\label{e:C3def}
\end{eqnarray}
which can (and does) tend to grow exponentially during numerical
evolutions. Second, this system does not satisfy the mathematical
condition (linear degeneracy) that prevents the formation of shocks
from smooth initial data~\cite{liu1979}.  The principal part of the
$\psi_{ti}$ component of Eq.~(\ref{e:AlviPsiDot}), for example, can be
written as $\partial_t N^i - N^k\partial_k N^i\simeq 0$; and these
terms have the same form as those responsible for shock formation in
the standard hydrodynamic equations.

We had previously developed ways to modify systems of this type to
eliminate either of these problems~\cite{Holst2004}.  However, these
methods produce systems that are not symmetric hyperbolic when both
problems are corrected simultaneously.  Here we present new
modifications that solve both problems without destroying
symmetric hyperbolicity.  We do this by adding appropriate multiples
of the constraint ${\cal C}_{iab}$ to each of the equations: $\gamma_1
N^i{\cal C}_{iab}$ to Eq.~(\ref{e:AlviPsiDot}), $\gamma_3 N^i{\cal
C}_{iab}$ to Eq.~(\ref{e:AlviPiDot}), and $\gamma_2 N {\cal C}_{iab}$
to Eq.~(\ref{e:AlviPhiDot}). These terms modify the principal parts of
the equations:
\begin{eqnarray}
\partial_t\psi_{ab}
-(1+\gamma_1)N^k\partial_k\psi_{ab} &\simeq& 0,
\label{e:NewPsiDot}\\
\partial_t\Pi_{ab} - N^k\partial_k\Pi_{ab} 
+ N g^{ki}\partial_k\Phi_{iab} -
\gamma_3 N^k\partial_k\psi_{ab}&\simeq &0,
\label{e:NewPiDot}\\ 
\partial_t\Phi_{iab}-N^k\partial_k\Phi_{iab}
+N\partial_i\Pi_{ab}-\gamma_2 N \partial_i\psi_{ab}&\simeq&0.
\label{e:NewPhiDot}
\end{eqnarray}
Choosing $\gamma_3=\gamma_1\gamma_2$ makes this new system symmetric
hyperbolic for any values of the parameters $\gamma_1$ and $\gamma_2$.
The symmetrizer metric (which defines the energy norm) for this new
system can be written as
\begin{eqnarray}
S_{\alpha\beta} du^\alpha du^\beta &=  m^{ab}m^{cd}\bigl(
&\Lambda^2 d\psi_{ac}d\psi_{bd}
+d\Pi_{ac}d\Pi_{bd}
\nonumber\\&&
-2\gamma_2d\psi_{ac}d\Pi_{bd}
+g^{ij}d\Phi_{iac}d\Phi_{jbd}\bigr),
\end{eqnarray}
where $m^{ab}$ is any positive definite metric (e.g.,
$m^{ab}=g^{ab}+t^a t^b$ or even $m^{ab}=\delta^{ab}$) and $\Lambda$
is a constant with dimension length${}^{-1}$.  This symmetrizer is positive
definite so long as $\Lambda^2>\gamma_2^2$.  

The eigenvectors of the characteristic matrix,
$n_kA^{k\,\alpha}{}_\beta$ (where $n_k$ is the outward directed unit
normal to the boundary of the computational domain), play an important
role in setting boundary conditions for first-order evolution systems.
Let $e^{\hat \alpha}{}_\beta$ denote the left eigenvectors with
eigenvalues $v_{(\hat\alpha)}$, defined by
\begin{eqnarray}
e^{\hat\alpha}{}_\mu n_k A^{k\,\mu}{}_\beta = v_{(\hat\alpha)}
e^{\hat\alpha}{}_\beta.\label{e:EigenvectorDef}
\end{eqnarray}
We use indices with hats (e.g., $\hat\alpha$) to label the
characteristic eigenvectors and eigenvalues, and $\hat\alpha$ is not
summed over in Eq.~(\ref{e:EigenvectorDef}).  The eigenvalues
$v_{(\hat\alpha)}$ are also called the characteristic speeds. The
characteristic matrices of symmetric hyperbolic systems have complete
sets of eigenvectors, so the matrix $e^{\hat\alpha}{}_\beta$ is
invertible in this case.  The characteristic fields, $u^{\hat\alpha}$,
are defined as the projections of the dynamical fields onto the
characteristic eigenvectors: $u^{\hat\alpha}\equiv
e^{\hat\alpha}{}_\beta u^\beta$.  Boundary conditions must be imposed
on each incoming characteristic field, i.e., each $u^{\hat\alpha}$
with negative characteristic speed,
$v_{(\hat\alpha)}<0$~\cite{Rauch1985, Secchi1996a, Secchi1996b}.  The
characteristic fields for the new GH evolution system,
Eqs.~(\ref{e:NewPsiDot})--(\ref{e:NewPhiDot}), are given by
\begin{eqnarray}
u^{\hat 0}_{ab}      &=& \psi_{ab},\\
u^{{\hat 1}\pm}_{ab} &=& \Pi_{ab} \pm n^i \Phi_{iab}
                           -\gamma_2 \psi_{ab}, \\
u^{\hat 2}_{iab}     &=& P_i{}^k \Phi_{kab},
\end{eqnarray}
\noindent
where $P_i{}^k=\delta_i{}^k - n_i n^k$.  The characteristic fields
$u^{\hat 0}_{ab}$ have coordinate characteristic speed
$-(1+\gamma_1)n_kN^k$, the fields $u^{{\hat 1}\pm}_{ab}$ have
speed $-n_kN^k\pm N$, and the fields
$u^{\hat 2}_{iab}$ have speed $-n_kN^k$.

The complete equations for our new first-order representation of the
GH evolution system (including all the non-principal
parts) are
\begin{eqnarray}
%PsiDot
\partial_t\psi_{ab}&-&(1+\gamma_1)N^k\partial_k\psi_{ab} 
 = - N\Pi_{ab}-\gamma_1N^i\Phi_{iab},
\label{e:psiEvol}\\
%PiDot
\partial_t\Pi_{ab} &-& N^k\partial_k\Pi_{ab} 
+ N g^{ki}\partial_k\Phi_{iab} - 
\gamma_1 \gamma_2 N^k \partial_k \psi_{ab}
\nonumber\\
&=&2N\psi^{cd}\bigl(  
  g^{ij} \Phi_{ica} \Phi_{jdb}
- \Pi_{ca} \Pi_{db}
- \psi^{ef}\Gamma_{ace}\Gamma_{bdf}
\bigr)
\nonumber\\&&
-2N\nabla_{(a}H_{b)}
- \half Nt^c t^d \Pi_{cd}\Pi_{ab}
-N t^c \Pi_{c i} g^{ij}\Phi_{jab}\nonumber\\
&&+N\gamma_0 \bigl[2\delta^c{}_{(a}t{}_{b)}-\psi_{ab}
t^c\bigr] ({H}_c+\Gamma_c)
- \gamma_1 \gamma_2 N^i \Phi_{iab},\label{e:PiEvol}\\
%PhiDot
\partial_t\Phi_{iab}&-&N^k\partial_k\Phi_{iab}
+N\partial_i\Pi_{ab}-N\gamma_2\partial_i\psi_{ab}
\nonumber\\
&=&\half N t^c t^d \Phi_{icd}\Pi_{ab}
+Ng^{jk}t^c\Phi_{ijc}\Phi_{kab}
-N\gamma_2\Phi_{iab}.\label{e:PhiEvol}
\end{eqnarray}
The terms on the right sides of
Eqs.~(\ref{e:psiEvol})--(\ref{e:PhiEvol}) are algebraic functions of
the dynamical fields.  The connection terms $\Gamma_{cab}$ appearing
on the right side of Eq.~(\ref{e:PiEvol}) are computed using
Eqs.~(\ref{e:Christoffel}), where it is understood that the partial
derivatives are to determined from the dynamical fields by
\begin{eqnarray}
\partial_t\psi_{ab}&=&-N\Pi_{ab}+N^i\Phi_{iab},
\label{e:psiDot}\\
\partial_i\psi_{ab}&=&\Phi_{iab}.\label{e:psiGrad}
\end{eqnarray}
Choosing the parameter $\gamma_2>0$ in this new system causes the
constraint ${\cal C}_{iab}$ to be exponentially
suppressed~\cite{Holst2004}, because the modified
Eq.~(\ref{e:PhiEvol}) implies an evolution equation for ${\cal
C}_{iab}$ having the form, $\partial_t{\cal C}_{iab}
-N^k\partial_k{\cal C}_{iab}\simeq -\gamma_2N{\cal C}_{iab}$.
Choosing $\gamma_1=-1$ makes the system
Eqs.~(\ref{e:psiEvol})--(\ref{e:PhiEvol}) linearly degenerate, which
implies that shocks do not form from smooth initial
data~\cite{liu1979}.  And choosing the parameter $\gamma_0>0$ causes
the constraint ${\cal C}_a$ to be exponentially suppressed, as
discussed in Sec.~\ref{s:ConstraintDamping}.

%%%%%%%%%%%%%%%%%%%%%%%%%%%%%%%%%%%%%%%%%%%%%%%%%%%%%%%%%%%%%%%%%%%%%%%%%%%%%%%
\section{Boundary Conditions}
\label{s:BoundaryConditions}
%%%%%%%%%%%%%%%%%%%%%%%%%%%%%%%%%%%%%%%%%%%%%%%%%%%%%%%%%%%%%%%%%%%%%%%%%%%%%%%

The modifications of the GH evolution system discussed in
Secs.~\ref{s:ConstraintDamping} and \ref{s:FirstOrderSystem} are
designed to damp out small constraint violations that may arise from
inexact initial data, numerical errors, etc.  These modifications will
do nothing, however, to prevent the influx of constraint violations
through the boundaries of the computational domain.  Constraint-preserving
boundary conditions are needed to prevent
this~\cite{Stewart1998,Calabrese2002a,Calabrese2003,Kidder2005,Sarbach2005}. 
Such boundary conditions can be formulated
once the propagation equations for the constraints are understood.  So
we derive a first-order system of evolution equations for the
constraints in Sec.~\ref{s:ConstraintEvolutionSystem}, use them to
derive constraint-preserving boundary conditions in
Sec.~\ref{s:ConstraintBC}, present boundary conditions for the
physical gravitational-wave degrees of freedom of the system in
Sec.~\ref{s:Physical-Boundary-Conditions}, and finally analyze the
well-posedness of the combined set of new boundary conditions in
Sec.~\ref{s:WellPosedness}.

\subsection{First-Order Constraint Evolution System}
\label{s:ConstraintEvolutionSystem}

The primary constraint of the GH system is the gauge constraint,
${\cal C}_a$, which we re-write here in terms of the first-order
dynamical fields:
\begin{eqnarray}
{\cal C}_a &=& H_a + g^{ij}\Phi_{ija}
		 +t^b \Pi_{ba}
                  -\half g_a^i \psi^{bc}\Phi_{ibc}
		 -\half t_a \psi^{bc}\Pi_{bc}.
\label{eq:OneIndexConstraint}
\end{eqnarray}
This expression differs from Eq.~(\ref{eq:GaugeConstraint}) only by
multiples of the constraint ${\cal C}_{iab}$.  In the following we use
this definition, Eq.~(\ref{eq:OneIndexConstraint}), rather than
Eq.~(\ref{eq:GaugeConstraint}), because it simplifies the form of the
constraint evolution system.  The evolution equation for ${\cal C}_a$,
Eq.~(\ref{e:ConstraintEvol}), is second order.  Thus, we must define
new constraint fields that represent the first derivatives of ${\cal
C}_a$ in order to reduce the constraint evolution system to
first-order form.  Thus we define new constraint fields ${\cal
F}_a$ and ${\cal C}_{ia}$ that satisfy
\begin{eqnarray}
{\cal F}_a    &\approx& t^c\partial_c {\cal C}_a
=N^{-1}(\partial_t {\cal C}_a
                            -N^i \partial_i {\cal C}_a),\\
{\cal C}_{ia} &\approx& \partial_i {\cal C}_a,
\end{eqnarray}
up to terms proportional to the constraints ${\cal C}_a$ and ${\cal
C}_{iab}$.  The following definitions of ${\cal F}_a$ and ${\cal
C}_{ia}$ accomplish this in a way that keeps the form of the
constraint evolution system as simple as possible:
\begin{eqnarray}
\label{eq:TimeDerivOfOneIndexConstraint}
{\cal F}_a &\equiv& 
\half g_a^i \psi^{bc}\partial_i \Pi_{bc}
- g^{ij} \partial_i \Pi_{ja}
- g^{ij} t^b \partial_i \Phi_{jba}
+ \half t_a \psi^{bc} g^{ij} \partial_i \Phi_{jbc}
\nonumber \\ &&
+ t_a g^{ij} \partial_i H_j 
+ g_a^i \Phi_{ijb} g^{jk}\Phi_{kcd} \psi^{bd} t^c
- \half g_a^i \Phi_{ijb} g^{jk}
  \Phi_{kcd} \psi^{cd} t^b
\nonumber \\ &&
- g_a^i t^b \partial_i H_b
+ g^{ij} \Phi_{icd} \Phi_{jba} \psi^{bc} t^d
- \half t_a g^{ij} g^{mn} \Phi_{imc} \Phi_{njd}\psi^{cd}
\nonumber \\ &&
- \fourth  t_a g^{ij}\Phi_{icd}\Phi_{jbe}
   \psi^{cb}\psi^{de}
+ \fourth  t_a \Pi_{cd} \Pi_{be} 
   \psi^{cb}\psi^{de}
- g^{ij} H_i \Pi_{ja}  
\nonumber \\ &&
- t^b g^{ij} \Pi_{b i} \Pi_{ja}
- \fourth  g_a^i \Phi_{icd} t^c t^d \Pi_{be}
  \psi^{be}
+ \half t_a \Pi_{cd} \Pi_{be}\psi^{ce}
  t^d t^b
\nonumber \\ &&
+ g_a^i \Phi_{icd} \Pi_{be} t^c t^b \psi^{de}
- g^{ij}\Phi_{iba} t^b \Pi_{je} t^e
- \half g^{ij}\Phi_{icd} t^c t^d \Pi_{ja}
\nonumber \\ &&
- g^{ij} H_i \Phi_{jba} t^b
+ g_{a}^i \Phi_{icd} H_b \psi^{bc} t^d
+\gamma_2\bigl(g^{id}{\cal C}_{ida}
-\half  g_a^i\psi^{cd}{\cal C}_{icd}\bigr)
\nonumber \\ &&
+ \half t_a \Pi_{cd}\psi^{cd} H_b t^b
- t_a g^{ij} \Phi_{ijc} H_d \psi^{cd}
+\half  t_a g^{ij} H_i \Phi_{jcd}\psi^{cd}
,
\end{eqnarray}
\begin{eqnarray}
\label{eq:TwoIndexConstraint}
{\cal C}_{ia} &\equiv& g^{jk}\partial_j \Phi_{ika} 
- \half g_a^j\psi^{cd}\partial_j \Phi_{icd} 
+ t^b \partial_i \Pi_{ba}
- \half t_a \psi^{cd}\partial_i\Pi_{cd}
\nonumber\\&&
+ \partial_i H_a 
+ \half g_a^j \Phi_{jcd} \Phi_{ief} 
\psi^{ce}\psi^{df}
+ \half g^{jk} \Phi_{jcd} \Phi_{ike} 
\psi^{cd}t^e t_a
\nonumber\\&&
- g^{jk}g^{mn}\Phi_{jma}\Phi_{ikn}
+ \half \Phi_{icd} \Pi_{be} t_a 
                            \left(\psi^{cb}\psi^{de}
                      +\half\psi^{be} t^c t^d\right)
\nonumber\\&&
- \Phi_{icd} \Pi_{ba} t^c \left(\psi^{bd}
                            +\half t^b t^d\right)
+ \half \gamma_2 \left(t_a \psi^{cd}
- 2 \delta^c_a t^d\right) {\cal C}_{icd}.
\end{eqnarray}
The remaining constraints needed to complete the GH constraint
evolution system are ${\cal C}_{iab}$ defined in Eq.~(\ref{e:C3def}),
and the closely related ${\cal C}_{ijab}$, defined by
\begin{eqnarray}
{\cal C}_{ijab} &=& 2\partial_{[i}\Phi_{j]ab}
= 2\partial_{[j}{\cal C}_{i]ab}.
\label{eq:FourIndexConstraint}
\end{eqnarray}

The complete collection of constraints for the GH
evolution system is therefore the set $c^A\equiv\{{\cal C}_{a}, {\cal
F}_{a}, {\cal C}_{ia}, {\cal C}_{iab}, {\cal C}_{ijab}\}$ defined in
Eqs.~(\ref{eq:OneIndexConstraint}),
(\ref{eq:TimeDerivOfOneIndexConstraint}),
(\ref{eq:TwoIndexConstraint}), (\ref{e:C3def}), and
(\ref{eq:FourIndexConstraint}).  (We use upper case Latin indices to
label the constraints.)  The constraints $c^A$ depend on the dynamical
fields $u^\alpha=\{\psi_{ab}, \Pi_{ab}, \Phi_{iab}\}$ and their
spatial derivatives $\partial_k u^\alpha$.  Thus the evolution of the
constraint fields $c^A$ is completely determined by the evolution of
the dynamical fields through
Eqs.~(\ref{e:psiEvol})--(\ref{e:PhiEvol}).  We have evaluated these
constraint evolution equations and have verified that they can be
written in the abstract form
\begin{eqnarray}
\partial_t c^A + A^{kA}{}_B(u)\partial_k c^B = F^A{}_B(u,\partial u)\, c^B,
\label{e:ConstraintEvolutionSystem}
\end{eqnarray}
where $A^{kA}{}_B$ and $F^A{}_B$ may depend on the dynamical fields
$u^\alpha$ and their spatial derivatives $\partial_k u^\alpha$.  Thus
the constraint evolution system closes: the time derivatives of the
constraints vanish initially when the constraints themselves vanish at
an initial time.  The principal part of the first-order constraint
evolution system turns out to be remarkably simple (given the
complexity of the expressions for the constraints themselves):
\begin{eqnarray}
\partial_t {\cal C}_a    &\simeq& 0,\\
\partial_t {\cal F}_a    &\simeq&  N^i \partial_i {\cal F}_a
                     + N g^{ij} \partial_i {\cal C}_{ja},\\
\partial_t {\cal C}_{ia} &\simeq& N^j \partial_j {\cal C}_{ia}
                     + N\partial_i {\cal F}_a,\\
\partial_t {\cal C}_{iab} &\simeq& 
             (1+\gamma_1) N^k \partial_k {\cal C}_{iab},    \\
\partial_t {\cal C}_{ijab} &\simeq& N^k \partial_k
                                    {\cal C}_{ijab}.       
\end{eqnarray}
This constraint evolution system is symmetric hyperbolic
with symmetrizer
\begin{eqnarray}
S_{AB} dc^A dc^B &=&
     m^{ab}\Bigl[d{\cal F}_ad{\cal F}_b 
     +g^{ij}\bigl(d{\cal C}_{ia}d{\cal C}_{jb}
     +g^{kl}m^{cd}d{\cal C}_{ikac}d{\cal C}_{jlbd}\bigr)
\nonumber\\&& 
     \qquad+ \Lambda^2\bigl(d{\cal C}_ad{\cal C}_b 
     +g^{ij}m^{cd}d{\cal C}_{iac}d{\cal C}_{jbd}\bigr)
\Bigr],
\end{eqnarray}
where $\Lambda^2$ is a positive constant and $m^{ab}$ is an
arbitrary positive definite metric.  The constraint energy for
this system is defined as
\begin{eqnarray}
{\cal E}_c = \int S_{AB} c^A c^B \sqrt{g} d^{\,3}x.
\label{e:ConstraintEnergy}
\end{eqnarray}
Since the constraint evolution system is hyperbolic, it follows (at
the continuum level) that the constraints will remain satisfied within
the domain of dependence of the initial data, if they are satisfied
initially.  

We have analyzed the solutions to this constraint evolution system for
the case of small constraint violations of solutions near flat space.
We find that all of the short-wavelength constraint violations are
damped at the rate $e^{-\gamma_0t}$, $e^{-\gamma_0t/2}$, or
$e^{-\gamma_2t}$.  So choosing $\gamma_0>0$ and $\gamma_2>0$ is
sufficient to guarantee that all of these constraints are suppressed.
This new first-order GH system therefore has the same constraint
suppression properties as the second-order system of Gundlach, et
al.~\cite{Gundlach2005} and Pretorius~\cite{Pretorius2005a}.

The constraint evolution system,
Eq.~(\ref{e:ConstraintEvolutionSystem}), is symmetric hyperbolic and
it will be useful to determine the characteristic constraint fields.
Thus, we evaluate the matrix of left eigenvectors of the constraint
evolution system $e^{\hat A}{}_B$ and their corresponding eigenvalues
$v_{(\hat A)}$ (or characteristic speeds).  The characteristic
constraint fields are defined (in analogy with the principal evolution
system) as the projections of the constraint fields onto these
eigenvectors: $c^{\hat A} \equiv e^{\hat A}{}_B c^B$.  The resulting
characteristic fields for this constraint evolution system are
\begin{eqnarray}
{c}^{{\hat 0}\pm}_a   &=&  {\cal F}_a  \mp n^k {\cal C}_{ka}, \\ 
{c}^{\hat 1}_{a}     &=& {\cal C}_a, \\
{c}^{\hat 2}_{ia}    &=& P^k{}_i {\cal C}_{ka}, \\
{c}^{\hat 3}_{iab} &=& {\cal C}_{iab}, \\
{c}^{\hat 4}_{ijab} &=& {\cal C}_{ijab}.
\end{eqnarray}
\noindent
The characteristic constraint fields $c^{{\hat 0}\pm}_a$ have
coordinate characteristic speeds $-n_lN^l\pm N$, the fields $c^{\hat
1}_a$ have speed $0$, the fields $c^{\hat 2}_{ia}$ and $c^{\hat
4}_{ijab}$ have speed $-n_lN^l$, and the fields $c^{\hat 3}_{iab}$
have speed $-(1+\gamma_1)n_lN^l$.  

%%%%%%%%%%%%%%%%%%%%%%%%%%%%%%%%%%%%%%%%%%%%%%%%%%%%%%%%%%%%%%%%%%%%%%%%%%%%%%
\subsection{Constraint-Preserving Boundary Conditions}
\label{s:ConstraintBC}

Boundary conditions must be imposed on all the incoming characteristic
fields $u^{\hat\alpha}$, i.e., all those with $v_{(\hat\alpha)}<0$ on
a particular boundary.  Thus, boundary conditions will typically be
needed for the characteristic field $u^{{\hat 1}-}_{ab}$, and
(depending on the value of the parameter $\gamma_1$ and the
orientation of the shift $N^k$ at the boundary) may also be needed for
$u^{\hat 0}_{ab}$ and/or $u^{\hat 2}_{iab}$.  Some of these boundary
conditions must be set by physical considerations, i.e., by specifying
what physical gravitational waves enter the computational domain.
Some of the boundary conditions can be used, however, to prevent the
influx of constraint violations.  This can be done by specifying the
incoming $u^{\hat\alpha}$ at the boundary in a way that ensures the
incoming characteristic constraint fields $c^{\hat A}$ also vanish
there. The incoming constraint fields for this system include
$c^{{\hat 0}-}_a$, and perhaps $c^{\hat 3}_{iab}$ and/or $c^{\hat
4}_{ikab}$ depending on $\gamma_1$ and $N^k$ at the boundary.  We find
that these incoming $c^{\hat A}$ are related to the incoming
$u^{\hat\alpha}$ by the following expressions:
\begin{eqnarray}
&&c^{{\hat 0}-}_a \approx
\sqrt{2}\Bigl[k^{(c} \psi^{d)}{}_a 
                        -\half k_a\psi^{cd}\Bigr]
                   d_\perp u^{{\hat 1}-}_{cd},\label{e:U1mPerp}
\\
&&n^i c^{\hat 3}_{iab} \approx  d_\perp u^{\hat 0}_{ab}, \label{e:U0Perp}\\
&&n^i c^{\hat 4}_{ikab} \approx  d_\perp u^{\hat 2}_{kab}.\label{e:U2Perp}
\end{eqnarray}
Here the notation $d_\perp u^{\hat\alpha}$ denotes the characteristic
projection of the normal derivatives of $u^{\hat\alpha}$ (i.e.,
$d_\perp u^{\hat\alpha}\equiv e^{\hat\alpha}_\beta
n^k\partial_ku^\beta$), and $\approx$ implies that algebraic terms and
terms involving tangential derivatives of the fields (i.e.,
$P^k{}_i\partial_ku^\alpha$) have not been displayed. The inward
directed null vector $k^c$ used here is defined as $k^c = (t^c -
n^c)/\sqrt{2}$.  The idea is to set the left sides of
Eqs.~(\ref{e:U1mPerp})--(\ref{e:U2Perp}) to zero to get Neumann-like
boundary conditions for the indicated components of $d_\perp
u^{\hat\alpha}$.  By imposing these conditions on $d_\perp
u^{\hat\alpha}$, we ensure that these incoming components of
$c^{\hat A}$ vanish.

We have found that a convenient way to impose boundary conditions of
this type is to set the incoming projections of the time derivatives
of $u^{\alpha}$, $d_t u^{\hat\alpha} \equiv e^{\hat\alpha}{}_\beta
\partial_t u^\beta$, in the following way:
\begin{eqnarray}
d_t u^{\hat\alpha} = D_t u^{\hat\alpha} + v_{(\hat\alpha)}
\bigl(d_\perp u^{\hat\alpha}-d_\perp u^{\hat\alpha}\bigr|_{BC}\bigr).
\label{e:BCMethod}
\end{eqnarray}
In this expression the terms $D_tu^{\hat\alpha}$ represent the
projections of the right sides of the evolution system,
Eqs.~(\ref{e:psiEvol})--(\ref{e:PhiEvol}); so the equations at
non-boundary points would simply be $d_t u^{\hat\alpha} =
D_tu^{\hat\alpha}$.  The term $d_\perp u^{\hat\alpha}\bigr|_{BC}$ is
the value to which $d_\perp u^{\hat\alpha}$ is to be fixed on the
boundary.  This form of the boundary condition replaces all of the
$d_\perp u^{\hat\alpha}$ that appears in $D_t u^{\hat\alpha}$ with
$d_\perp u^{\hat\alpha}\bigr|_{BC}$.  Applying this method to
the constraint-preserving boundary conditions in
Eqs.~(\ref{e:U1mPerp})--(\ref{e:U2Perp}), we obtain the following
rather simple conditions
\begin{eqnarray}
     d_t u^{\hat 0}_{ab}    &=& D_t u^{\hat 0}_{ab} 
                             - (1+\gamma_1)n_jN^jn^k c^{\hat 3}_{kab},
                               \label{e:V1BC} 
\\
d_t u^{{\hat 1}-}_{ab} &=& 
      \bigl[\half P_{ab}P^{cd} 
	- 2 l_{(a}P_{b)}{}^{(c}k^{d)} 
	+ l_a l_b k^c k^d\bigr]
      D_t u^{{\hat 1}-}_{cd} 
\nonumber\\&& 
      +\sqrt{2}(N+n_jN^j)
      \bigl[
	l_{(a} P_{b)}{}^{c} 
	-\half P_{ab}l^c 
	-\half l_a l_b k^c
	\bigr]c^{{\hat 0}-}_{c},\label{e:U1mBC}
\\
     d_t u^{\hat 2}_{kab}   &=& D_t u^{\hat 2}_{kab} 
                             - n_l N^l n^i P^j{}_k c^{\hat 4}_{ijab}.  
                               \label{e:Z1BC}
\end{eqnarray}
The quantity $P_{ab}$ in these expressions is the
projection tensor, $P_{ab} = \psi_{ab} + t_a t_b - n_a n_b$, and
the outgoing null vector $l^a$ is defined by $l^a = (t^a + n^a)/\sqrt{2}$.
 
%%%%%%%%%%%%%%%%%%%%%%%%%%%%%%%%%%%%%%%%%%%%%%%%%%%%%%%%%%%%%%%%%%%%%%%%%%%%%%%
\subsection{Physical Boundary Conditions}
\label{s:Physical-Boundary-Conditions}

The constraint-preserving boundary conditions presented in
Eqs.~(\ref{e:V1BC})--(\ref{e:Z1BC}) restrict only four degrees of
freedom of $u^{{\hat 1}-}_{ab}$.  Two of the remaining degrees of
freedom represent the physical gravitational waves, and the final four
represent gauge freedom.  We choose to characterize and control this
gravitational wave freedom in terms of the incoming parts of the Weyl
curvature.  The propagating components of the Weyl tensor can be
written as
\begin{eqnarray}
w^{\pm}_{ab} 
&=& \bigl(P_a{}^c P_b{}^d - \half P_{ab}
P^{cd}\bigr)\bigl(t^e\mp n^e\bigr)\bigl(t^f\mp n^f\bigr)
C_{cedf}.
\end{eqnarray}
We showed in Ref.~\cite{Kidder2005} that these components of the Weyl
tensor are the incoming and outgoing (respectively) characteristic
fields of the curvature evolution system that follows from the Bianchi
identities.  The $w^{\pm}_{ab}$ are proportional to the Newman-Penrose
curvature spinor components $\Psi_4$ (outgoing) and $\Psi_0$ (ingoing)
respectively.  We also note that the spatial components of
$w^{\pm}_{ij}$ are equal to the components of the Weyl tensor
characteristic fields $2U^{8\pm}_{ij}$ defined in our paper on
constraint-preserving boundary conditions for the KST
system~\cite{Kidder2005}.  
The expression for the Weyl tensor in terms of our first order
variables is unique only up to terms proportional to constraints; it is
possible to choose these constraint terms so that the $w^{\pm}_{ij}$ depend on
the normal derivatives of $u^{\hat\alpha}$ in the following way:
\begin{eqnarray}
w^{\pm}_{ab} &\approx&
\bigl(P_a{}^c P_b{}^d - \half P_{ab}P^{cd}\bigr)
\bigl(d_\perp u^{{\hat 1}\pm}_{cd}+\gamma_2 d_\perp u^{\hat 0}_{cd}\bigr).
\label{eq:WeylCharFieldsInTermsOfNormalDerivs}
\end{eqnarray}
Thus a physical boundary condition can be placed on the relevant
components of $u^{{\hat 1}-}_{ab}$ using the method of
Eq.~(\ref{e:BCMethod}) by setting
\begin{eqnarray}
d_t u^{{\hat 1}-}_{ab} &= 
\bigl(P_a{}^c P_b{}^d &- \half P_{ab}P^{cd}\bigr)\times
\nonumber\\&&
\bigl[D_t u^{{\hat 1}-}_{cd} 
-(N+n_jN^j)(w^{-}_{cd}-\gamma_2 n^ic^{\hat 3}_{icd})\bigr].
\label{e:PhysicalBC}
\end{eqnarray}
We can also inject incoming physical gravitational waves with a
predetermined waveform $\dot h{}_{ab}(t,x)$ through the boundary of
the computational domain by setting
\begin{eqnarray}
d_t u^{{\hat 1}-}_{ab} &=& \bigl(P_a{}^c P_b{}^d - \half
P_{ab}P^{cd}\bigr) \dot h{}_{cd}(t,x).
\label{e:InjectionBC}
\end{eqnarray}
The case $\dot h{}_{ab}=0$ corresponds to an isolated system with no
incoming gravitational waves. 

More generally we can combine the constraint-preserving, physical
no-incoming radiation, and the injected gravitational wave boundary
conditions by setting $d_t u^{{\hat 1}-}_{ab}$ equal to the sum of the
right sides of Eqs.~(\ref{e:U1mBC}), (\ref{e:PhysicalBC}), and
(\ref{e:InjectionBC}), and setting the time derivatives of the other
incoming fields according to Eqs.~(\ref{e:V1BC}) and~(\ref{e:Z1BC}).
Note that this set of combined boundary conditions holds the {\it pure
gauge} components of $u^{{\hat 1}-}_{ab}$ constant in time; other
boundary conditions on the gauge degrees of freedom are of course
possible but are not considered here.

%%%%%%%%%%%%%%%%%%%%%%%%%%%%%%%%%%%%%%%%%%%%%%%%%%%%%%%%%%%%%%%%%%%%%%%%%%%%%%%
\subsection{Well-posedness}
\label{s:WellPosedness}

The well-posedness of the initial-boundary value problem can be
analyzed using the Fourier-Laplace technique~\cite{Gustafsson1995}. We
have applied this method to the GH system with the combined set of
boundary conditions presented here: we treat the case of
high-frequency perturbations of flat spacetime in a slicing with flat
spatial metric, unit lapse, and a constant shift that is tangent to
the boundary.  Applying the Fourier-Laplace technique to this case
yields a necessary (but not sufficient) condition for well-posedness,
the so-called determinant condition~\cite{Gustafsson1995}; failure to
satisfy this condition would mean the system admits exponentially
growing solutions with arbitrarily large growth rates.  We have
verified that this determinant condition is satisfied for the GH
system using the combined set of boundary conditions presented here.

%%%%%%%%%%%%%%%%%%%%%%%%%%%%%%%%%%%%%%%%%%%%%%%%%%%%%%%%%%%%%%%%%%%%%%%%%%%%%%%
\section{Numerical Results}
\label{s:NumericalResults}

%--------------------BEGIN--FIGURE-----------------------------------------
\begin{figure} 
\begin{center}
\hbox{
\includegraphics[width=2.45in]{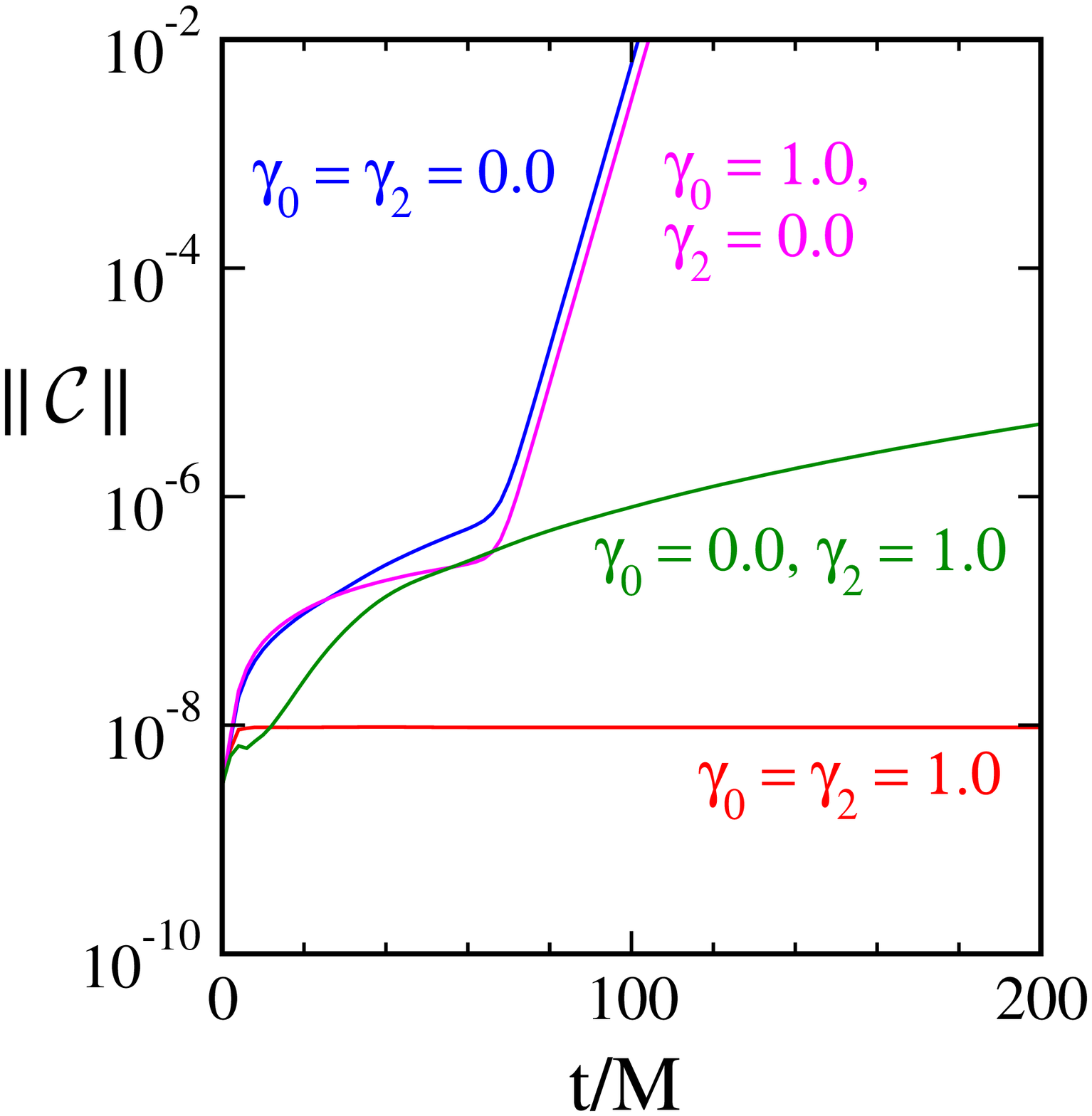}
\hspace{0.1in}
\includegraphics[width=2.45in]{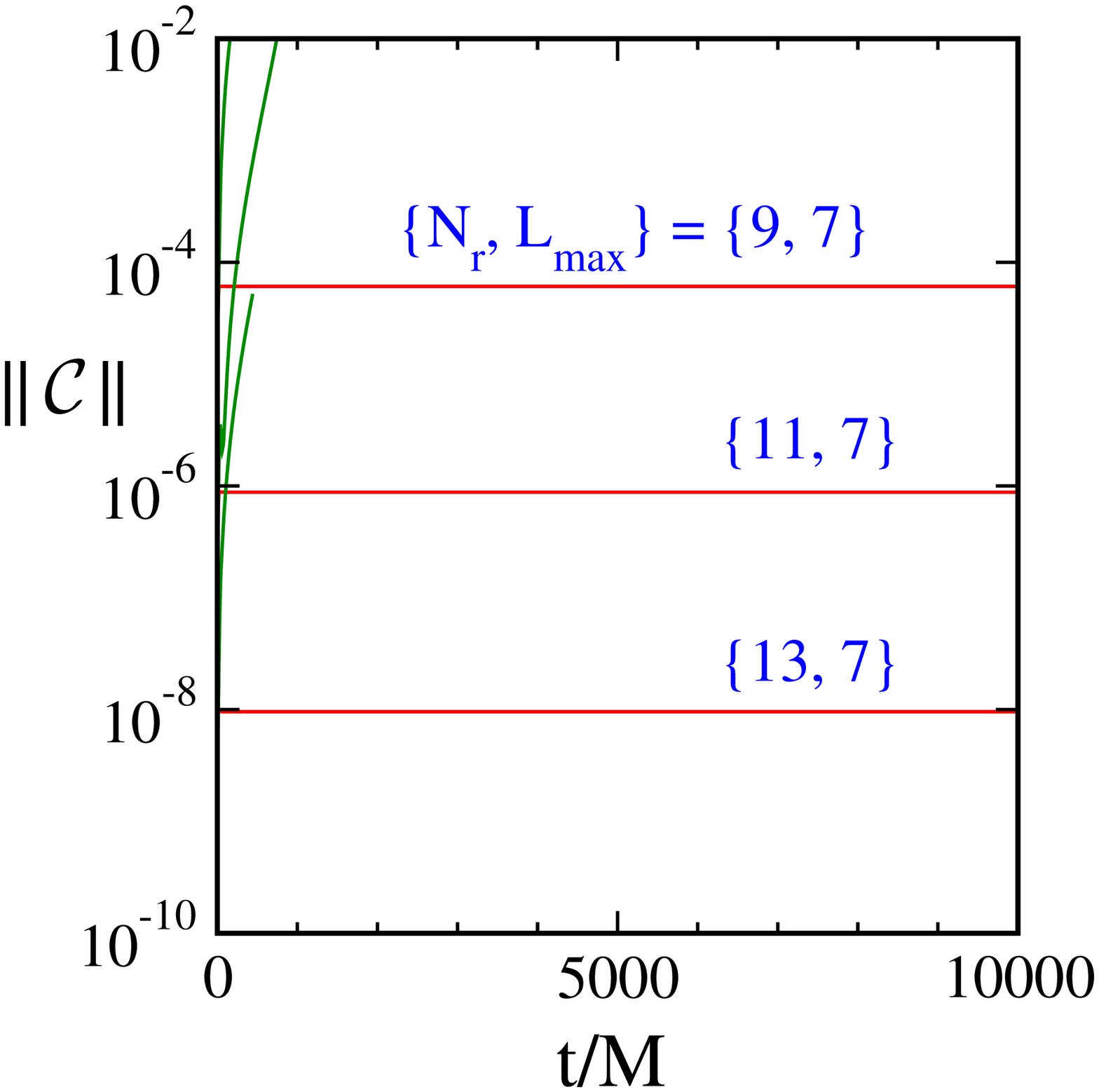}
}
\end{center}
\vspace{-0.25cm}
\caption{Evolution of constraint violations for Schwarzschild initial
data.  Left figure shows evolutions using various values of the
constraint damping parameters $\gamma_0$ and $\gamma_2$ using
numerical resolution $\{N_r,L_{max}\}=\{13,7\}$.  Right figure shows
the long timescale evolution of the same data for three different
numerical resolutions.
\label{f:fig1}}
\vspace{-0.25cm}
\end{figure}
%---------------------END--FIGURE-----------------------------------------
In this section we describe several numerical tests of the new
first-order GH evolution system.  First we test the effectiveness of
the two constraint damping terms included in
Eqs.~(\ref{e:psiEvol})--(\ref{e:PhiEvol}) by evolving Schwarzschild
initial data (in Kerr-Schild coordinates).  These tests are performed
on a computational domain consisting of a spherical shell that extends
from $r_{\mathrm min}=1.8M$ (just inside the event horizon) to
$r_{\mathrm max}=11.8M$, where $M$ is the mass of the black hole.  In
these evolutions we ``freeze'' the values of the incoming
characteristic fields to their initial values by setting $d_t u^{\hat
\alpha}=0$ on the boundaries for all incoming fields (i.e., all
$u^{\hat\alpha}$ with $v_{(\hat\alpha)}<0$).  We performed these
numerical evolutions using spectral methods as described for example
in Ref.~\cite{Kidder2005} for a range of numerical resolutions
specified by the parameters $N_r$ (the highest radial spectral basis
function) and $L_{max}$ (the highest spherical-harmonic basis
function).  Figure~\ref{f:fig1} illustrates the results of these tests
for several values of the constraint damping parameters $\gamma_0$ and
$\gamma_2$.  These tests show that without constraint damping the
extended GH evolution system is extremely unstable, but with
constraint damping the evolutions of the Schwarzschild spacetime are
completely stable up to $t=10,000M$ (and forever, we
presume).  These tests also illustrate that both the $\gamma_0$ and
the $\gamma_2$ constraint damping terms are essential for stable
evolutions.

Constraint violations in Fig.~\ref{f:fig1} (and in the rest of this
paper) are measured with the constraint energy ${\cal E}_c$ defined in
Eq.~(\ref{e:ConstraintEnergy}).  Since ${\cal E}_c$ is not
dimensionless, its magnitude has no absolute meaning.  We construct an
appropriate scale with which to compare ${\cal E}_c$ by evaluating the
$L^2$ norm of the spatial gradients of the dynamical fields,
\begin{eqnarray}
||\partial u||^2 &= \int g^{ij}m^{ab}m^{cd}\Bigl(&
\Lambda^2\partial_i\psi_{ac}\partial_j\psi_{bd}
+\partial_i\Pi_{ac}\partial_j\Pi_{bd}
\nonumber\\&&
+g^{kl}\partial_i\Phi_{kac}\partial_j\Phi_{lbd}\Bigr)\sqrt{g}d^3x.
\end{eqnarray}
The dimensionless constraint norm $||{\cal C}||$ shown in these
figures is defined as
\begin{eqnarray}
||{\cal C}||=\frac{\sqrt{{\cal E}_c}}{||\partial u||},
\end{eqnarray}
which is a meaningful measure of the relative size of constraint
violations in a particular solution.  In the figures shown here we
evaluate $||{\cal C}||$ with $m^{ab}=\delta^{ab}$ and the dimensional
constant $\Lambda=1/M$.

%--------------------BEGIN--FIGURE-----------------------------------------
\begin{figure} 
\begin{center}
\includegraphics[width=2.45in]{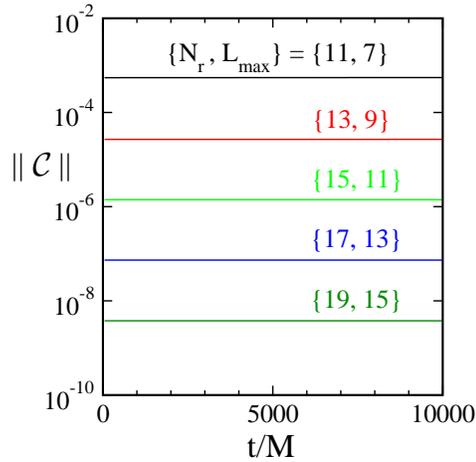}
\end{center}
\caption{Evolution of constraint violations for Kerr initial data
with spin parameter $\vec a = (0.1, 0.2, 0.3)$ for several numerical 
resolutions.
\label{f:fig2}}
\end{figure}
%---------------------END--FIGURE-----------------------------------------
Our second numerical test evolves the somewhat more challenging
initial data for a Kerr black hole (in Kerr-Schild coordinates) on a
computational domain consisting of a spherical shell that extends from
$r_{\mathrm min}=1.8M$ (just inside the event horizon) to $r_{\mathrm
max}=21.8M$.  We use two subdomains, each having numerical resolution
$\{N_r,L_{max}\}$, to cover this region.  The spin of the Kerr
spacetime used here is $\vec a = (0.1, 0.2, 0.3)M$, where the
magnitude of this vector determines the Kerr spin parameter $a=|\vec
a|\approx 0.374 M$, and the direction determines the orientation of
the Kerr rotation axis relative to the quasi-Cartesian coordinate
system used in our code.  For this test we use the combined set of physical
and constraint-preserving boundary
conditions discussed at the end of Sec.~\ref{s:Physical-Boundary-Conditions}.
Figure~\ref{f:fig2} shows that numerical
evolutions of this Kerr spacetime are stable and numerically
convergent to $t=10,000M$ (and forever, we presume) using a range of
numerical resolutions.

\begin{figure} 
\begin{center}
\hbox{
\includegraphics[width=2.45in]{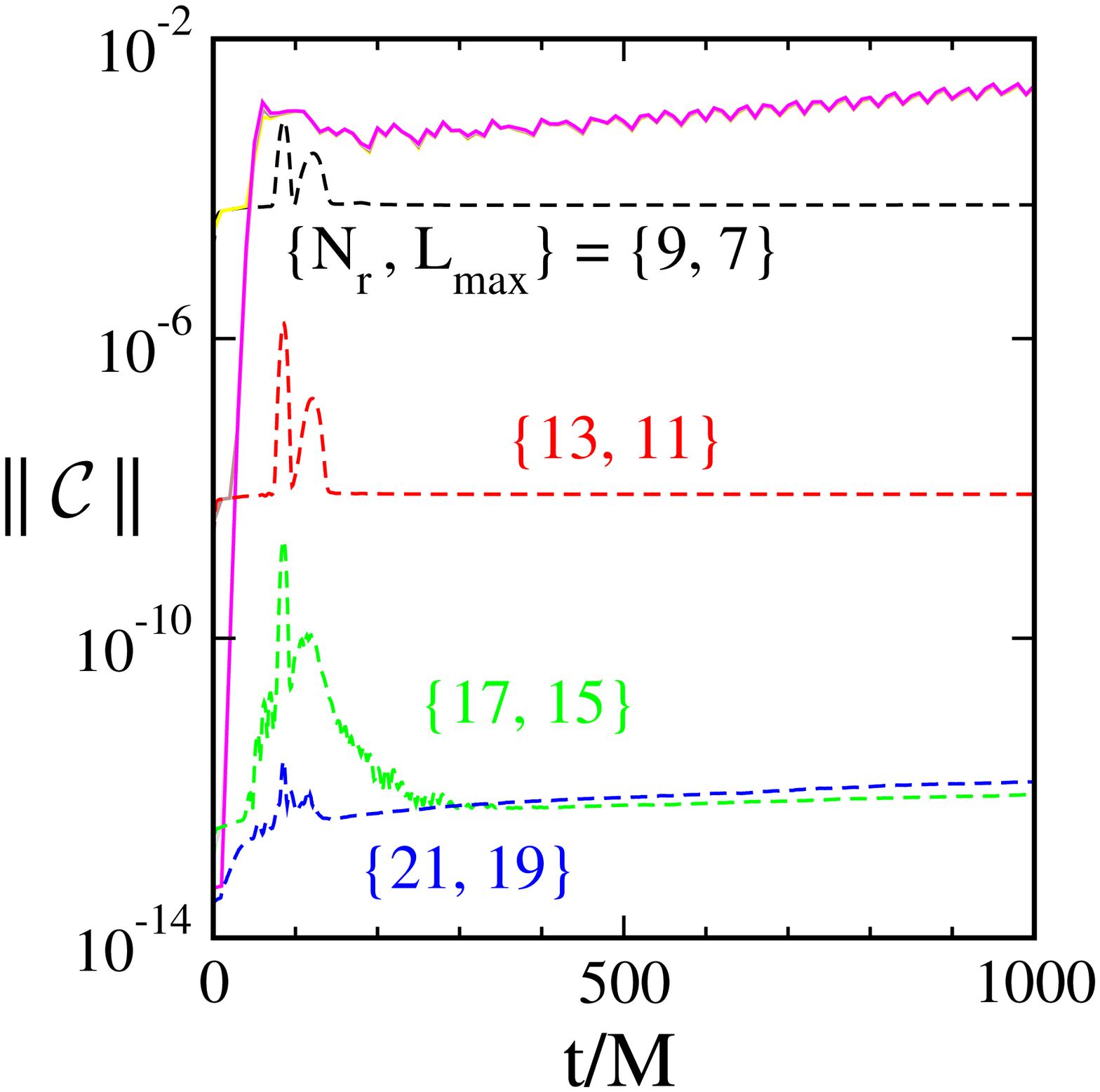}
\hspace{0.1in}
\includegraphics[width=2.45in]{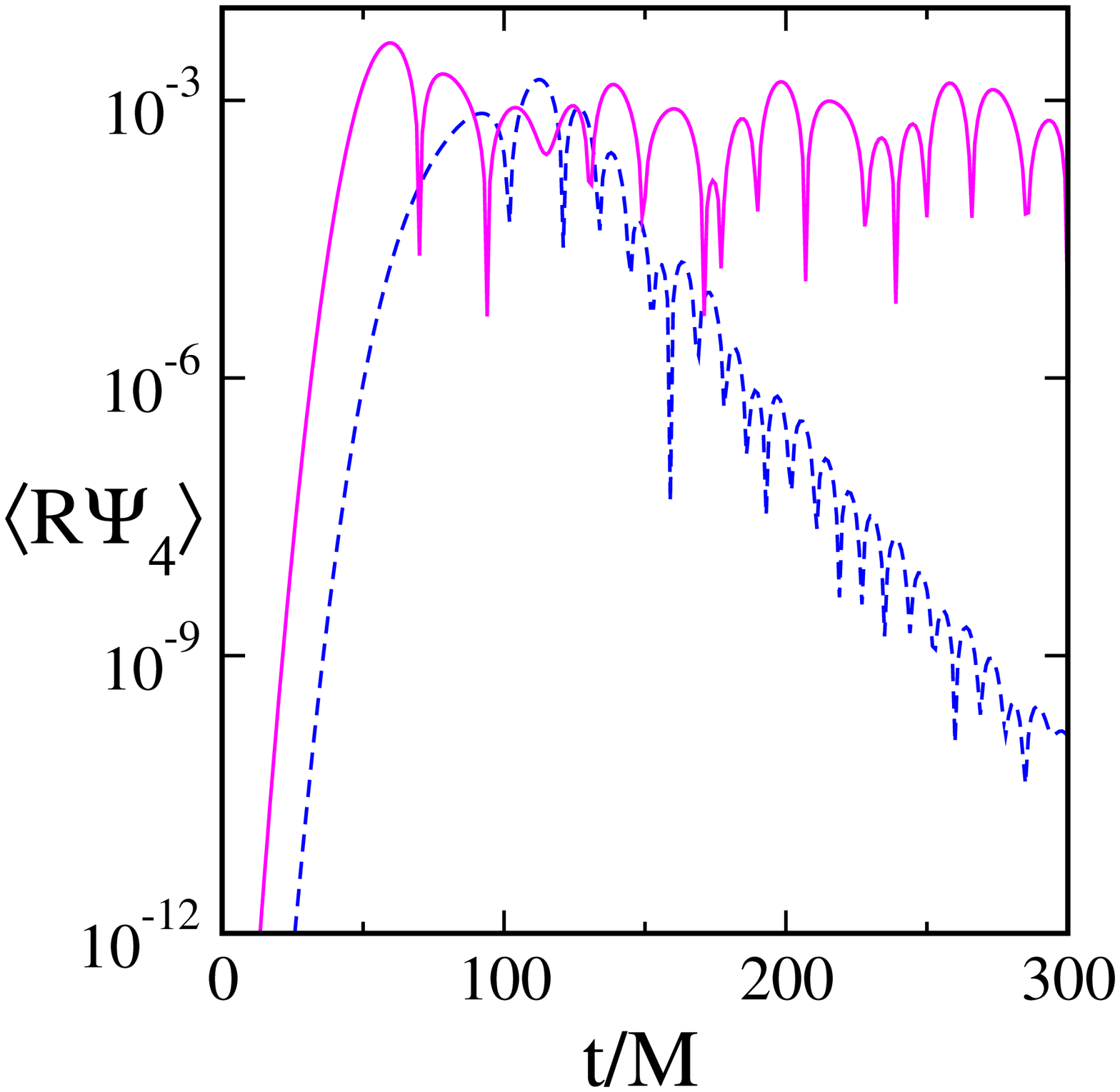}
}
\end{center}
\vspace{-0.25cm}
\caption{Evolution of Schwarzschild initial data perturbed by a
gravitational wave pulse with amplitude $10^{-3}$.  Left figure depicts
constraint violations at various numerical resolutions, and the right
figure shows $\Psi_4$ averaged over the outer boundary of the
computational domain at a single numerical resolution.  Solid curves
use freezing boundary conditions and dashed curves use constraint-preserving
and physical boundary conditions.
\label{f:fig3}}
\end{figure}

Our third numerical test is designed to demonstrate the effectiveness
of our new constraint-preserving boundary conditions.  This test
consists of evolving a black-hole spacetime perturbed by an incoming
gravitational wave pulse.  We start with Schwarzschild initial data,
and perturb it via the incoming gravitational wave boundary condition
described in Eq.~(\ref{e:InjectionBC}) with $\dot h_{ab} = \dot f(t)
(\hat x^a \hat x^b + \hat y^a \hat y^b - 2 \hat z^a \hat z^b)$ where
$\hat x^a$, $\hat y^a$, and $\hat z^a$ are the components of the
coordinate basis vectors, $\hat x^a\partial_a =\partial_x$, etc.  For
these evolutions we use an incoming gravitational wave pulse whose
time profile is $f(t)={\cal A}\, e^{-(t-t_p)^2/{ w}^2}$ with ${\cal
A}=10^{-3}$, $t_p=60M$, and ${w}=10M$.  This test is performed on the
same computational domain described above for the second numerical test.
Figure~\ref{f:fig3} illustrates the results of these tests using
two types of boundary conditions: frozen-incoming-field 
(i.e., $d_t u^{\hat\alpha}=0$ for $v_{(\hat\alpha)}<0$) 
boundary conditions (solid curves) and the new combined set of
constraint-preserving and physical boundary conditions discussed at the
end of Sec.~\ref{s:Physical-Boundary-Conditions} (dashed curves).  The graph
on the left in Fig.~\ref{f:fig3} shows that constraint violations
converge toward zero as the numerical resolution is increased when
the new boundary conditions are used, but not when
frozen-incoming-field boundary conditions are used.

%--------------------BEGIN--FIGURE-----------------------------------------
%---------------------END--FIGURE-----------------------------------------
The graph on the right in Fig.~\ref{f:fig3} shows the outgoing
physical gravitational wave flux (measured on the outer boundary of
the computational domain) computed using frozen-incoming-fields (sold
curve) and the new constraint-preserving and physical
(dashed curve) boundary conditions.
These evolutions were computed with numerical resolution
$\{N_r,L_{\max}\} =\{21,19\}$.  We measure the outgoing gravitational
wave flux with the quantity $\langle R\Psi_4\rangle$, which is the
Weyl curvature component $\Psi_4$ averaged over the outer boundary of
our computational domain:
\begin{eqnarray}
4\pi \langle R \Psi_4 \rangle^2 &=& \int |\Psi_4|^2 d^{\,2} V.
\end{eqnarray}
Here $4\pi R^2$ is the proper surface area of the boundary, and
$d^{\,2}V$ represents the proper area element on this boundary.  Since
$\Psi_4$ falls off like $1/R$, this quantity should be independent of
$R$ (asymptotically).  The dashed curve on the right in
Fig.~\ref{f:fig3} clearly shows quasi-normal oscillations with
frequency $\omega M = 0.376-0.089\,i$ (determined by a numerical fit
to these data).  This is in good agreement with the frequency of the
most slowly damped quasi-normal mode of the black hole: $\omega M =
0.37367-0.08896\,i$~\cite{chandra75}.  It is interesting to note that
the solid curve---using frozen-incoming-fields boundary
conditions---gives qualitatively incorrect results for the physical
gravitational waveform, even though the level of constraint violations
is fairly small numerically in this case.  This is not surprising because
the magnitude of constraint violations in this case is comparable
to the size of the injected gravitational wave pulse.

%%%%%%%%%%%%%%%%%%%%%%%%%%%%%%%%%%%%%%%%%%%%%%%%%%%%%%%%%%%%%%%%%%%%%%%%%%%%%%%
%Acknowledgment
\ack We thank Steven Detweiler, Yvonne Choquet-Bruhat, Harald
Pfeiffer, Frans Pretorius, Olivier Sarbach, Tilman Sauer, Saul
Teukolsky, and James York for helpful discussions concerning this
work.  L.L. thanks the Isaac Newton Institute for Mathematical Sciences
for their hospitality during a visit in which a portion of this work was
completed.  This work was supported in part by a grant from the Sherman
Fairchild Foundation to Caltech and Cornell, by NSF grants
PHY-0099568, PHY-0244906 and NASA grants NAG5-10707, NAG5-12834 at
Caltech, and by NSF grants PHY-0312072, PHY-0354631, and NASA grant
NNG05GG51G at Cornell. Some of the computations for this project were
performed with the Tungsten cluster at the National Center for
Supercomputing Applications.

%%%%%%%%%%%%%%%%%%%%%%%%%%%%%%%%%%%%%%%%%%%%%%%%%%%%%%%%%%%%%%%%%%%%%%%%%%%%%%%
\section*{References}
%%%%%%%%%%%%%%%%%%%%%%%%%%%%%%%%%%%%%%%%%%%%%%%%%%%%%%%%%%%%%%%%%%%%%%%%%%%%%%%
\bibliographystyle{unsrtabbrv} 
\bibliography{References}
%\begin{thebibliography}{99}
%\end{thebibliography}

\end{document}